# “Animal Spirits” in America, April 2009

Elliott Middleton[*]

April 8, 2009

**Abstract**

“Adaptation level and ‘animal spirits’” (Middleton 1996) presented a psychophysical theory of confidence levels based on the oldest and probably most widely observed law in psychology, the sensitivity to adaptation level. For Americans, whose attachments to employment and livelihood are often tenuous in a country without a European-style social safety net, it is the sensitivity to the unemployment rate that drives confidence levels. In “’Animal spirits’ and recession forecasting” (Middleton 2001; see also Ball 2001), the adaptation level theoretic metric of “animal spirits,” *A,* was combined with the slope of the U.S. Treasury yield curve in a logistic recession forecasting model that has correctly predicted every turning point in the economy since then. The model currently forecasts increasing confidence and an end to the recession in mid-2009. The question, given the severity of the current slump globally, is whether this forecast is plausible in the face of possibly very large increases in macroeconomic volatility.

The model proposed for “animal spirits” in Middleton (1996) was:

$$A_t = -\frac{U_t - \overline{U}_t}{\sigma_t} \qquad (1)$$

where $U_t$ is the unemployment rate, $\overline{U}_t$ is a 48-month rolling exponential average and $\sigma_t$ is the sample standard deviation of the 48-month rolling sample, and

[*] email: elliott.middleton@aya.yale.edu . The author is a senior risk manager in industry. This publication is personal.

$$\overline{U}_t = \frac{1}{\sum_{\ell=0}^{47} e^{-0.03\ell}} \sum_{\ell=0}^{47} e^{-0.03\ell} U_{t-\ell} \;.$$

(2)

The figure shows the very close relationship of $A$ to the Michigan Survey of Consumer Sentiment (green; shading represents recessions; blue is a forecast of "animal spirits"):

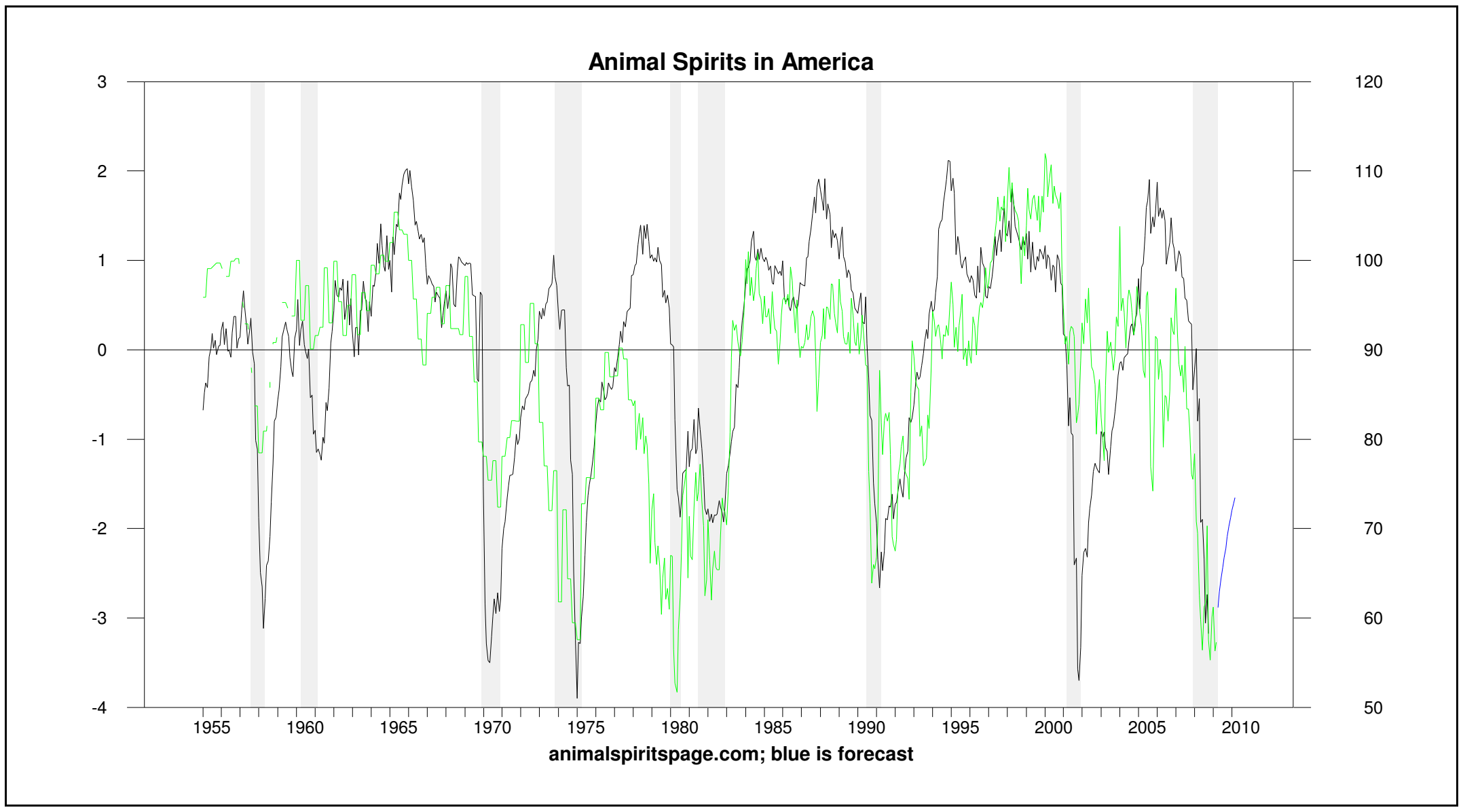

Some interpretative observations: the stock market crash of 1987 took place near a peak of confidence, and hence was not followed by a recession; the 9-11 attacks on the World Trade Center occurred at a time of already very low confidence; confidence levels during the current slump are at generational lows comparable only to the 1973-1974 and 1980-1982 slumps. Note that when $A < 0$, "animal spirits" are literally "negative," while $A > 0$ corresponds to periods when survey measures are elevated.

There is a behavioral theory of the business cycle implied by the plunging of the $A$ metric into recessions. As the unemployment rate approaches a "natural" or otherwise minimal rate in an expansion, confidence is naturally squeezed as its trailing adaptation level approaches the current unemployment rate. At some point in every cycle, unemployment ceases to go down, the adaptation level catches up, confidence evaporates, and confidence-dependent propensities to spend for consumption or investment flag. Unemployment rises above the adaptation level and the "animal spirits" become depressed. A critical state or tipping point is triggered in which falling confidence depresses demand and then output, unemployment goes up, confidence goes down, and an "animal spirits" avalanche occurs. The response of spending propensities to $A$ may be nonlinear and stronger in the neighborhood of $A = 0$ approached from above, much in the manner of the slope of a utility function becoming steeper below a reference point (Kahneman 1979). "Animal spirits" might be viewed as psychological "wealth."

Happily, rising unemployment does not generate limitless despair. If agents scale their perceptions of $U_t$ by the volatility measure, they will be less adversely affected by it. And the rising adaptation level will also cause a revision of perceptions; confidence falls only while $U_t$ is rising faster than $\overline{U}_t$. At some point, businesses decide that they have cut enough and $U_t$ reaches a local maximum. One implication of this theory is that excessive macroeconomic stabilization will cause agents' reactions to turbulence to be more violent than given a higher level of variation. This is not an argument for cyclical unemployment without welfare provisions, but for a level of macroeconomic volatility that "keeps agents on their toes" (Middleton 1996 discusses this in the context of the Wundt curve) and ready to meet adaptive challenges.

The next graph shows the volatility measure, which is forecast to skyrocket in the U.S. over the coming year. The forecast is based on the unemployment rate climbing to about 11 percent, which is more pessimistic than the consensus forecast now. This level of macroeconomic volatility has not been seen over the postwar period, and implies a dramatic rescaling of perceptions. However, increasing volatility, *cet. par.*, implies increasing confidence and perhaps a better ability to function amid turbulence.

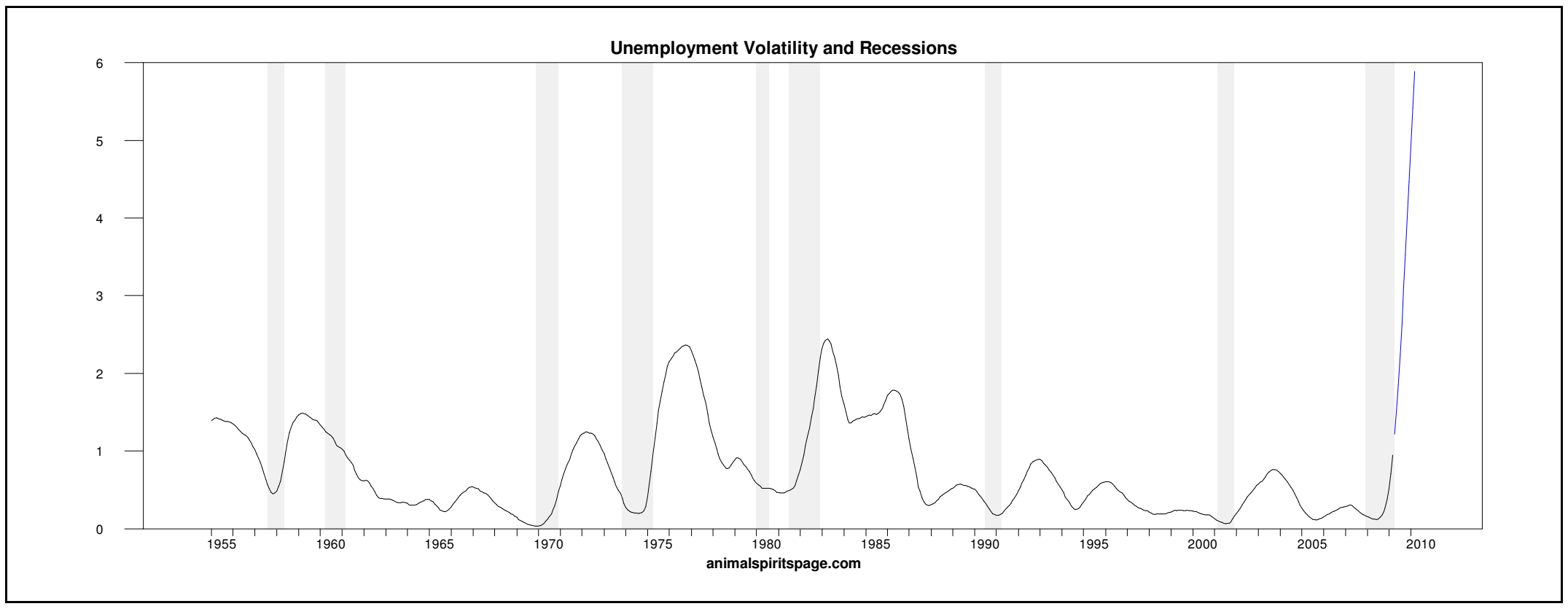

The following graph shows the unemployment rate and its adaptation level as of March 2009 (green is forecast). Note that the unemployment rate has risen after the last two recessions ended:

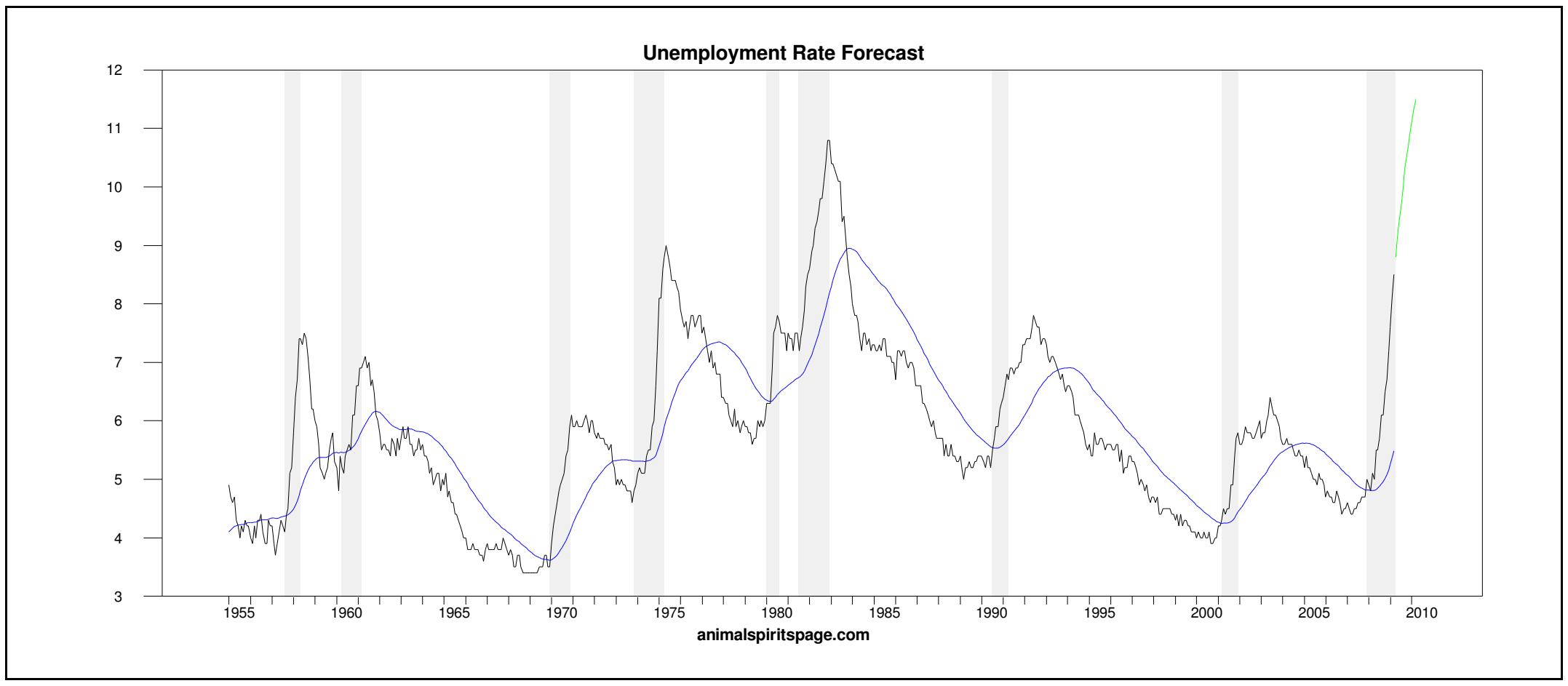

Recall from the first graph that even given the continuing climb in $U_t$ that is forecast, confidence recovers in about mid-2009. In fact, both *A* and the Michigan survey measure have increased slightly from their generational lows.

Bottoms in "animal spirits" correspond closely to bottoms of Industrial Production and to tops of risk premia as measured by Moody's Baa – Aaa ratings:

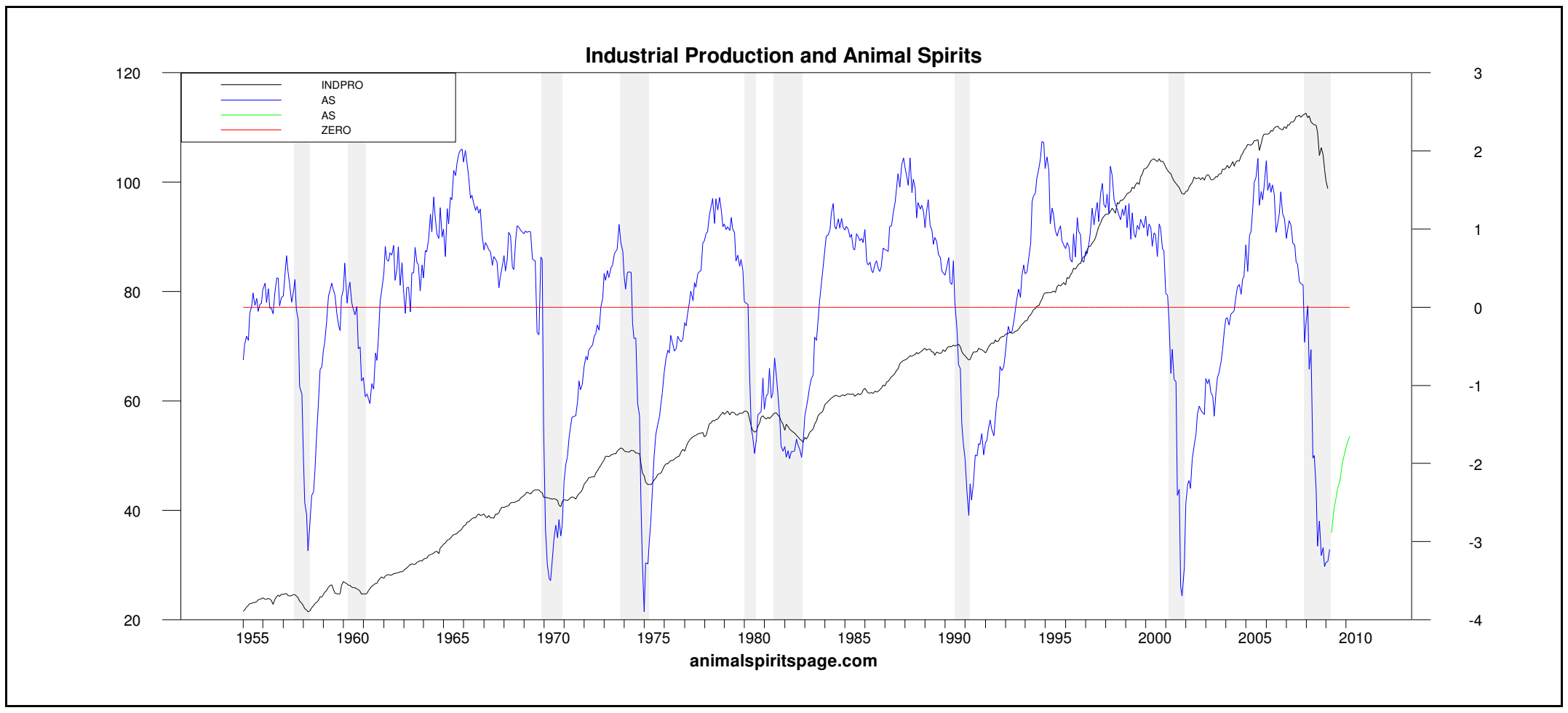

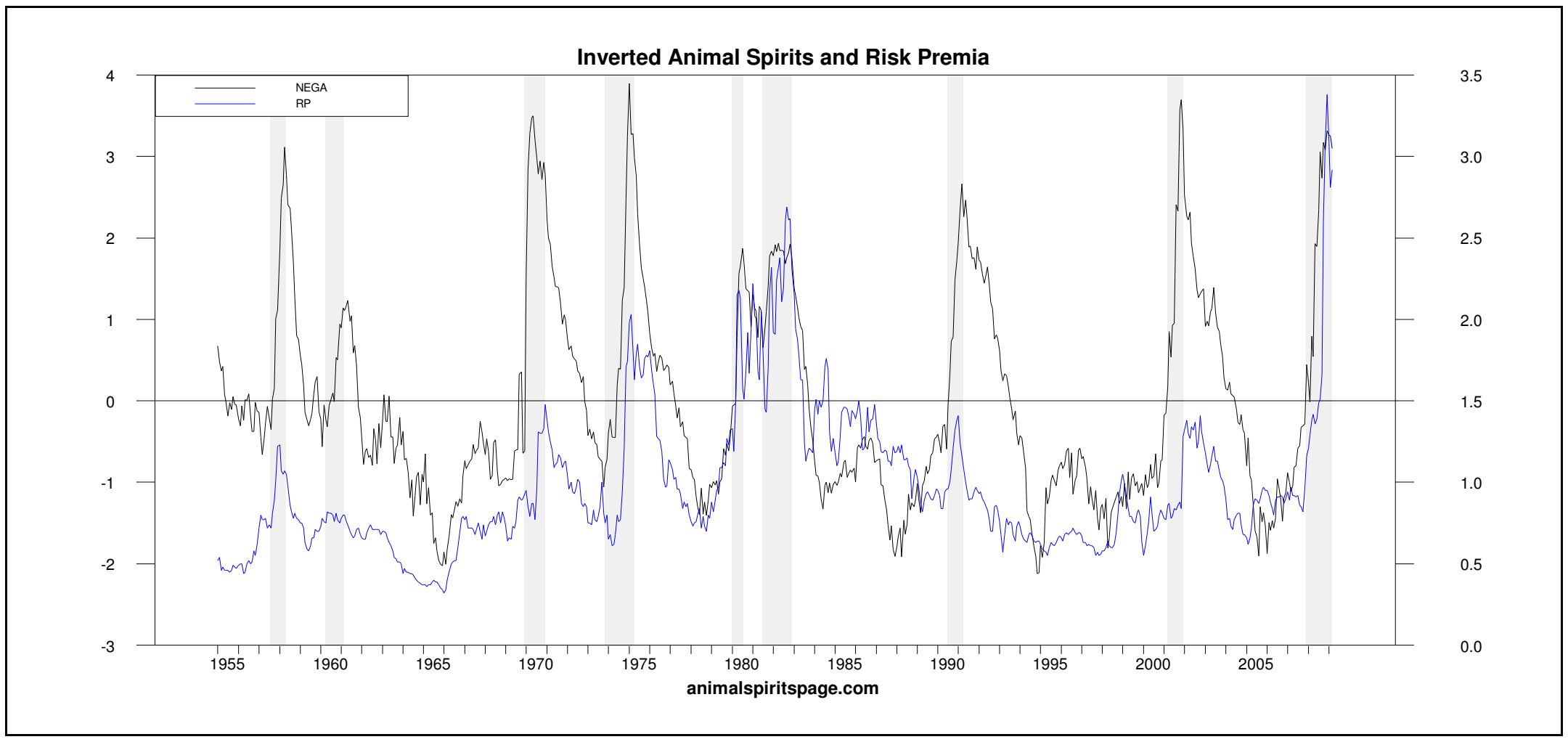

Investment spending as a percent of GDP—the context of Keynes' original remarks on "animal spirits" (Keynes 1936)—also track the *A* metric (pink) closely, especially at bottoms:

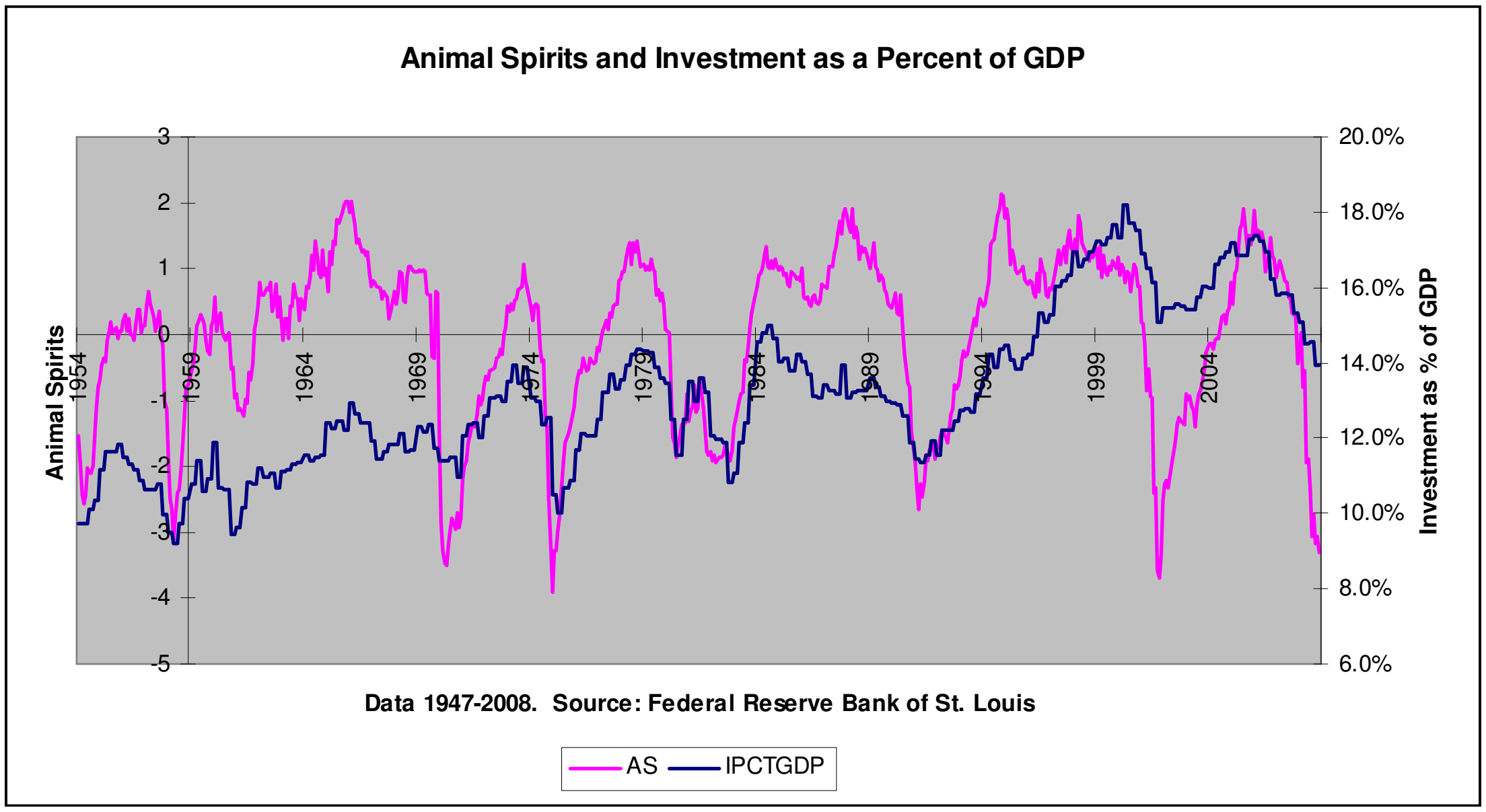

Psychophysical sensitivities to macroeconomic variables are similar to the sensitivities of many animals to the availability of foodstuffs in their environments. A measure similar to that in equation (1) but minus the initial minus sign can be constructed on real GDP with very similar results:

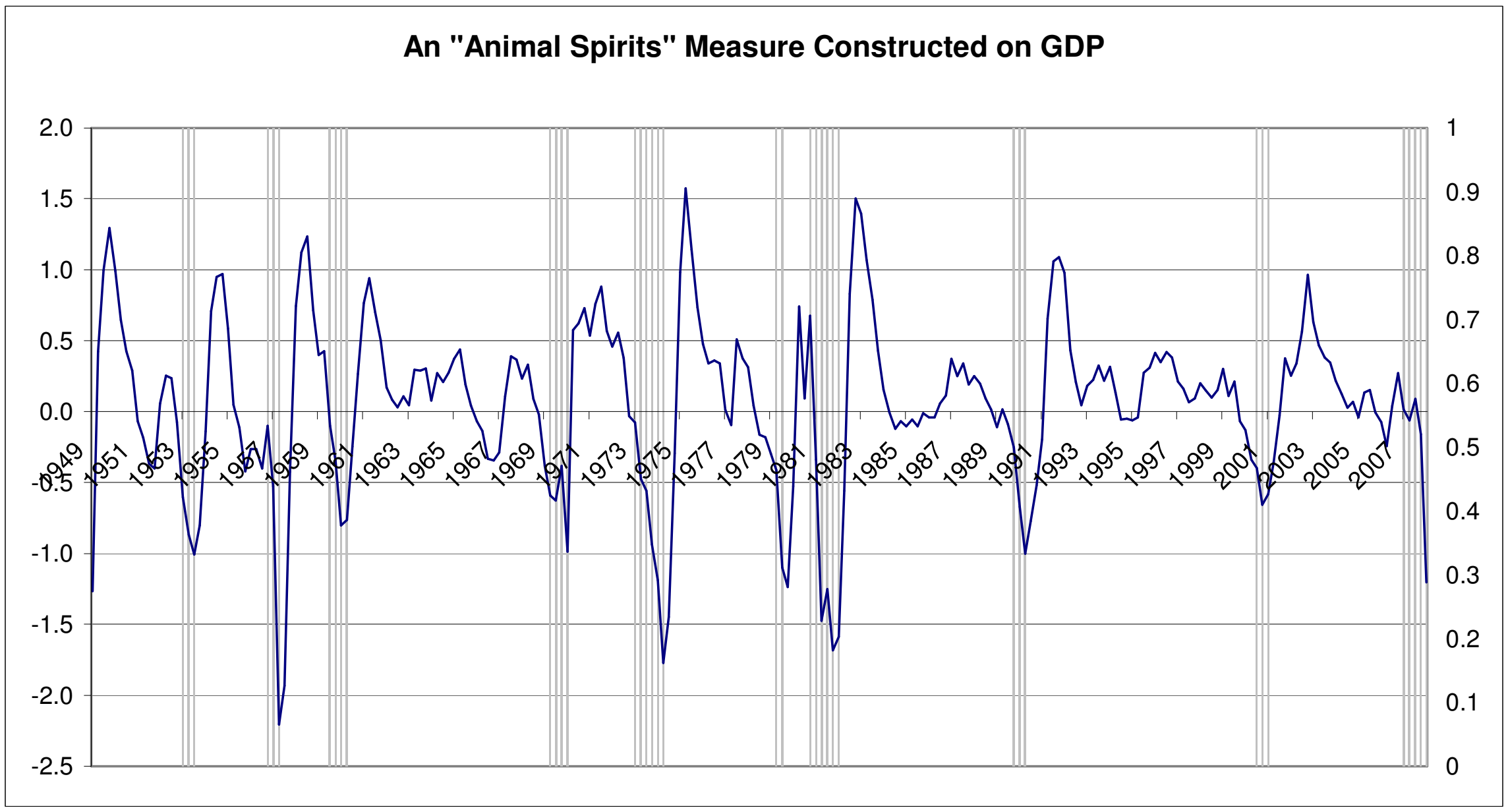

When the *A* metric is combined with the slope of the constant maturity Treasury yield curve from 1-year to 10-years, a venerable forecasting tool in its own right, in a logistic regression forecasting model (Middleton 2001) the result has proven to be accurate in a qualitative way over the current decade out of sample (Ball 2001):

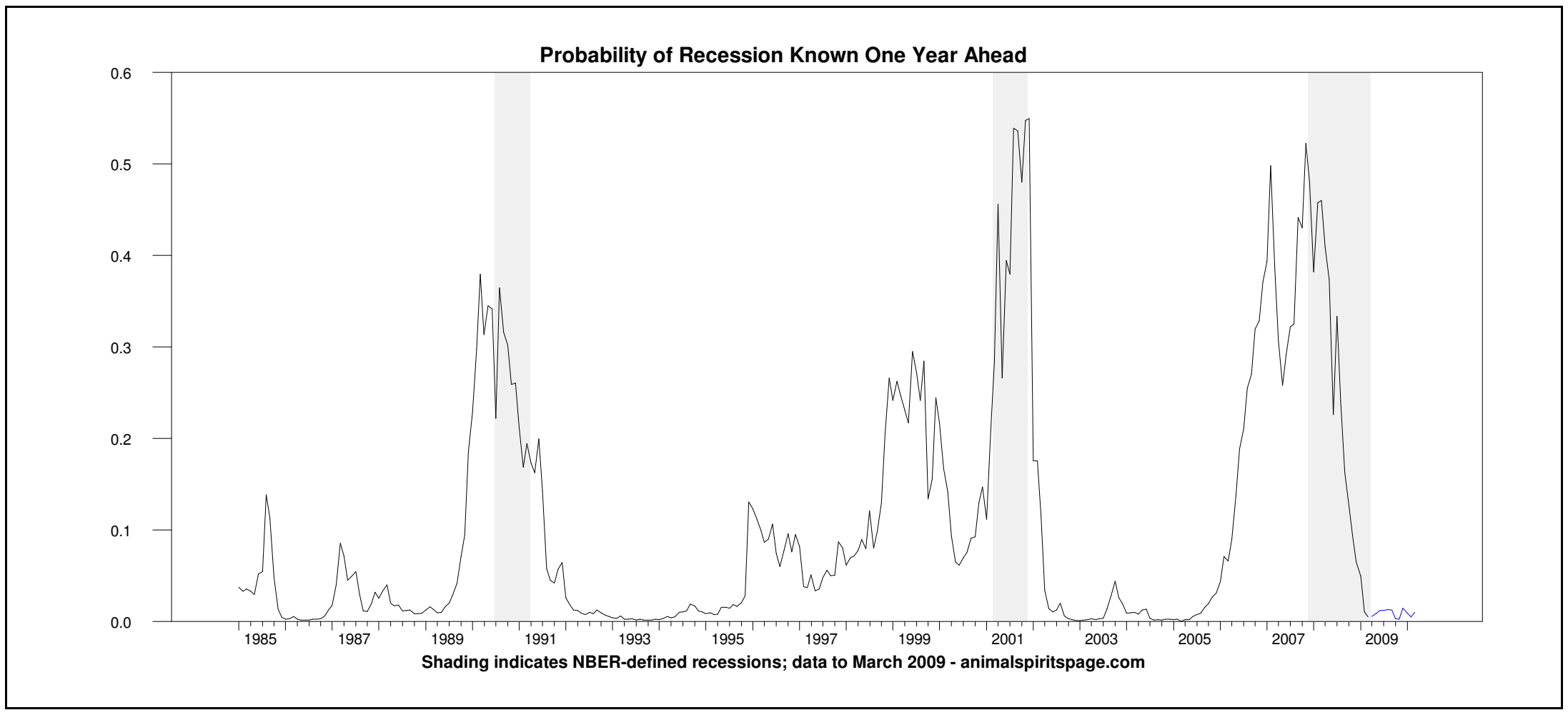

The model uses 12-month lags on the slope of the yield curve and the *A* metric; no exogenous variables are forecast. The yield curve tends to invert about a year before recessions begin, an effect widely attributed to bond market expectations of lower long-term rates in the future. As noted in Middleton (2001) there are many variants of yield curve forecasting models, but most have some structural instability. The *A* metric reduced that problem. The graph for the whole sample is based on a single equation estimated on data available before 1990:

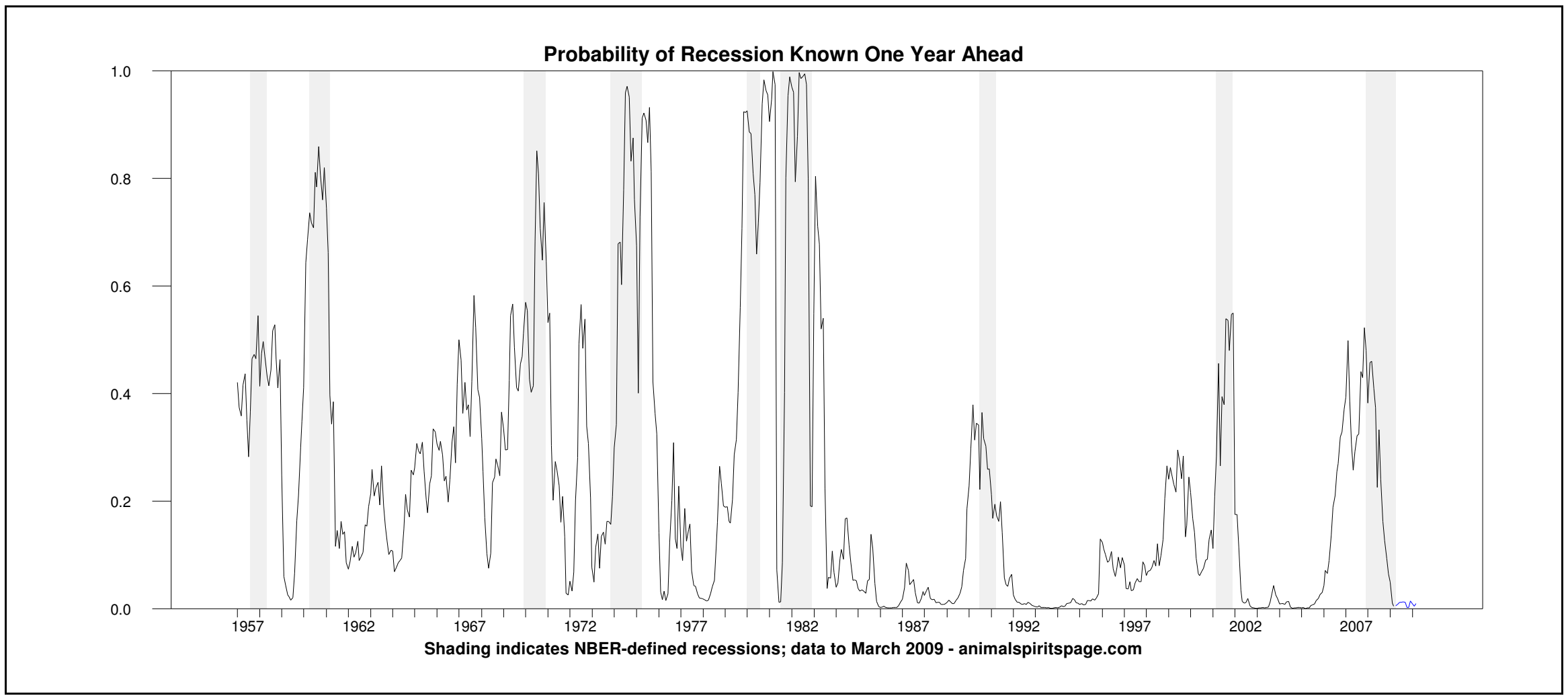

The recession "probability" falls to near zero in early 2009, consistent with the revival of "animal spirits" implicit in the unemployment rate forecast.

## Discussion

This paper provides a novel means of forecasting American "animal spirits" or confidence levels given a plausible unemployment rate forecast, and suggests a behavioral theory of the business cycle based on the principle of sensitivity to adaptation level observed throughout the animal domain. Americans are especially sensitive to the unemployment rate; similar measures based on output may be more closely related to confidence levels in societies with a social safety net. The paper suggests that investment responds to "animal spirits" as Keynes believed. The *A* metric is useful in conjunction with the slope of the yield curve in a logistic regression forecasting model that performs extremely well in forecasting turning points, something that the economics profession has not been particularly good at. In the context of the current global business slump the present model offers some hope that confidence levels, output and investment will rise in the near term, even if unemployment increases, although "animal spirits" will remain depressed.

A number of uncertainties underlie the model.

First, the measurement of the unemployment rate has changed over time in ways that tend to reduce the widely reported unemployment rate (Williams 2009). If "discouraged" and "working part-time for economic reasons" segments were included in the widely reported rate, it would rise from the March 2009 value of 8.5 percent to 15.6 percent of the civilian labor force. The "birth-death model" that attempts to estimate self-employment of laid-off workers is widely believed to be a fiction during recessions; this adjustment added 140,000 jobs in March or about -0.1 percent of unemployment rate (Bureau of Labor Statistics 2009). Finally, if agents begin to doubt the veracity of published statistics, it might result in a further loss of confidence and a deterioration of the modeled relationships.

Second, given the global nature of the current slump and its tremendous momentum to the downside, is the forecast credible? VoxEu (Eichengreen and O'Rourke 2009) find that the current slump has accelerated to the downside faster than the Great Depression. Put another way, even if "animal spirits" improve somewhat as Americans and others adapt to the new realities ("been down so long it looks like up to me"), will there be sufficient effective demand (disposable income and credit) among consumers and businesses to support increased levels of output? It only takes marginal changes in consumption propensities to change aggregate demand, and the recent spike in the consumer saving rate might be reversed somewhat. Also, there will be fiscal stimulus money available through slight individual income tax cuts in the relevant time frame that may boost demand. Possibly offsetting is the negative wealth effect from the collapse of housing and stock prices. The logistic recession forecasting model, incorporating information from the yield curve, is forecasting recovery. As the Chinese say, *yin* always gives way to *yang*, sooner or later. Americans will not stay depressed forever.

Third, unprecedented levels of labor market volatility may induce a deeper sense of uncertainty and loss of confidence instead of rescaling perceptions as proposed. In this regard, the model may be misspecified, as very high levels of volatility might amplify rather than mitigate losses of confidence. If the denominator of $A$ is set equal to one and we use the given (more pessimistic than consensus) unemployment rate forecast, confidence plunges over the coming year to a new postwar low:

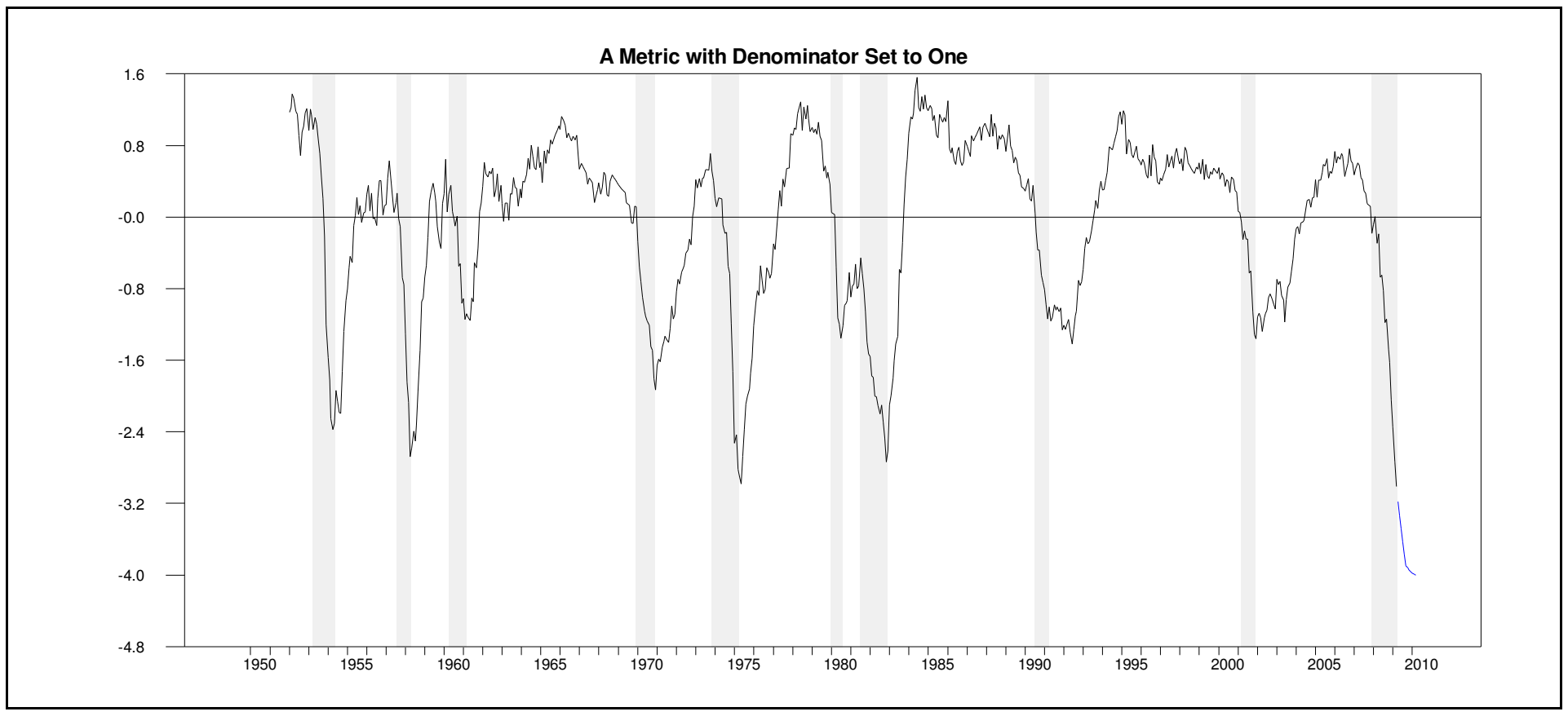

Finally, there is a widespread perception in America that the social contract is broken, a factor that is exogenous to the present model. A former chief economist of the International Monetary Fund has described America as resembling an emerging nation, crushed by debts, whose government treasury is being looted by a tiny elite, a nation on its way to becoming a failed state (Johnson 2009). A long wave theory of Anglo-American history and a prescient prophesy of the past decade in America written in the mid-1990s (Strauss and Howe 1997) suggests that Anglo-America follows the pulse of a historical *saeculum* with crises every ~80 years, the sequence to date being 1485, 1588, 1689, 1781, 1863, 1944. The authors describe each crisis as an *ekpyrosis*, a death and rebirth of the republic, as the old social contract is destroyed and a new one fashioned, often during a major war. H.E. Stanley (2008) notes in comments on the current crisis that network theory and physical phenomena have shown that sparsely connected networks are more stable than densely connected ones, the implication being that with attempts at global coordination of policy responses to the crisis, the fluctuations associated with the crisis are likely to become more severe than without such actions. In the United States, monetary and fiscal policy dials are set at levels of stimulation never before encountered simultaneously. The adaptive response to volatility of many types in the coming period will be a critical determinant of "animal spirits" and of outcomes generally.

Further research from an adaptation level theoretic perspective would appear to be promising. To the extent that "animal spirits" are determined by a fairly stable unconscious process, which is characteristic of adaptation level effects in many areas of physiological psychology (Helson 1964), the approach can be expected to yield powerful results. The ability to forecast "animal spirits" may enable better forecasts of other variables, as the *A* metric and similar variables become part of larger macroeconomic models.